\documentclass[twocolumn,twoside,slac_two]{revtex4}
\usepackage{graphicx}
\usepackage{fancyhdr}
\begin{document}

\title{Theoretical Review on CP Violation in Rare $B$ decays}

\author{Cheng-Wei Chiang}
\affiliation{Department of Physics and Center for Mathematics and Theoretical
  Physics, National Central University, Chungli, Taiwan 320, R.O.C.}
\affiliation{Institute of Physics, Academia Sinica, Taipei, Taiwan 115, R.O.C.}

\begin{abstract}
  We discuss several issues related to direct CP violation in rare $B$ meson
  decays.  We review the use of CP asymmetries in extracting information of
  strong and weak phases, how the experimental data fit into the overall
  picture, and the current status of the $K \pi$ puzzle.  We also examine the
  flavor symmetry assumption using closely related decay modes and extract the
  weak phase $\gamma$ from certain $B \to K^* \pi$ and $\rho K$ decays.
\end{abstract}

\maketitle

\thispagestyle{fancy}

\section{Importance of CP violation}

As pointed out by Sakharov \cite{Sakharov:1967dj} in the 60's, one of the
necessary conditions for the observed Universe is CP violation in physical
processes.  In the standard model (SM) of particle physics, the only source of
CP violation is given by the so-called Kobayashi-Maskawa mechanism
\cite{Kobayashi:1973fv} in the quark sector.  In weak transitions, the up-type
quarks and the down-type quarks are coupled through the $3 \times 3$
Cabibbo-Kobayashi-Maskawa (CKM) matrix \cite{Cabibbo:1963yz,Kobayashi:1973fv},
which contains a CP-violating phase.  Therefore, studying and understanding the
origin of such a phase in the SM is crucial to particle physics and cosmology
\cite{Hou:2008xd}.  More importantly, it is possible to shed some light on new
physics in such studies.

Due to its hierarchical structure, the CKM matrix has one useful unitarity
condition, which connects its first and third column:
\begin{eqnarray}
V_{ud} V_{ub}^* + V_{cd} V_{cb}^* + V_{td} V_{tb}^* = 0 ~.
\end{eqnarray}
This relation has a special status because it renders on a complex plane a
triangle that has all sides about the same size (and so are the angles).  An
important program of current $B$-factories is to use various processes to
overconstrain this unitarity triangle (UT).  Through such an exercise, we hope
not only to measure precisely the sides and angles of the UT but also to obtain
hints of physics beyond the SM that provides additional CP-violating sources.

The indirect CP violation in the $B$ system has been first established in the
charmonium modes in 2001, and is now measured at a precision better than $5\%$.
Soon after that first measurement, the direct CP violation in the $B$ system has
also been observed in the $B^0 \to K^+ \pi^-$ decay mode in 2004.  This is a
result of the interference between color-allowed tree and QCD penguin
amplitudes.  In the following, we concentrate exclusively on the direct CP
asymmetries in rare $B$ decays.

\section{Direct CP asymmetries in rare $B$ decays}

Among all processes, charmless two-body hadronic $B$ decays are often sensitive
to the $V_{td}$ and $V_{ub}$ matrix elements that involve CP-violating phases
through $B$-$\bar B$ mixing and/or decay amplitudes.  With increasing precision
on their branching ratios and CP asymmetries, these rare decay modes provide
additional useful constraints on the UT.

Consider the decay of a $B$ meson into some final state $f$ and its
CP-conjugated one.  Assuming the process involves two amplitudes, one has
\begin{eqnarray}
A(B \to f) &=&
   A_1 + A_2 e^{i (\phi + \delta)} ~,
\nonumber \\
A({\overline B} \to {\overline f}) &=&
   A_1 + A_2 e^{i (-\phi + \delta)} ~,
\label{eq:amps}
\end{eqnarray}
and
\begin{eqnarray}
{\cal A}_{CP}
&=& \frac{\Gamma({\bar B} \to {\bar f}) - \Gamma(B \to f)}
         {\Gamma({\bar B} \to {\bar f}) + \Gamma(B \to f)}
\nonumber \\
&=& \frac{2 A_1 A_2 \sin\phi \, \sin\delta}
       {A_1^2 + A_2^2 + 2 A_1 A_2 \cos\phi \cos\delta} ~.
\label{eq:ACP}
\end{eqnarray}
In Eqs.~(\ref{eq:amps}) and (\ref{eq:ACP}), $\phi$ and $\delta$ denote
respectively the relative strong and weak phases between the two amplitudes.
Therefore, observing a sizable CP asymmetry in the decays requires the
interference of at least two amplitudes with large relative strong and weak
phases.

Currently, information of direct CP asymmetries in hadronic $B$ decays is
collected by the BaBar, Belle, CLEO, CDF and D{\O} Collaborations.  In
Table~\ref{tab:DCPV}, we list the asymmetries that deviate from zero at more
than $3\sigma$ level.

\begin{table}[h]
\caption{Direct CP asymmetries of rare $B$ decays measured at the $3\sigma$
  level or more, all quoted from Ref.~\cite{HFAG}.
  \label{tab:DCPV}}
\begin{tabular}{lll}
\hline\hline
DCP & Value & Level \\
\hline
${\cal A}_{CP}(K^+ \pi^-)$ & $-0.097 \pm 0.012$ & $8.1 \sigma$ \\
${\cal A}_{CP}(\pi^+ \pi^-)$ & $0.38 \pm 0.07$ & $5.4 \sigma$ \\
${\cal A}_{CP}(K^{*0} \eta)$ & $0.19 \pm 0.05$ & $3.8 \sigma$ \\
${\cal A}_{CP}(\rho^0 K^+)$ & $0.37 \pm 0.11$ & $3.4 \sigma$ \\
${\cal A}_{CP}(\rho^{\pm} \pi^{\mp})$ & $-0.13 \pm 0.04$ & $3.3 \sigma$ \\
${\cal A}_{CP}(\eta K^+)$ & $-0.27 \pm 0.09$ & $3 \sigma$ \\
\hline\hline
\end{tabular}
\end{table}

The program of studying CP-violating phases is partly impeded by the lack of
full dynamical understanding in hadronic physics, including both strong phases
and hadronic matrix elements.  A lot of progress in perturbative approaches
\cite{Beneke-Lu-Rothstein} has been made in recent years.  However, it is still
a challenging problem to obtain from first principles sufficiently large strong
phases, as required by some of the observed large CP asymmetries.

An alternative approach employs the flavor SU(3) symmetry to help relating
parameters in amplitudes with the same flavor topology.  Strangeness-conserving
($\Delta S = 0$) amplitudes mediating $b \to q {\bar q} d$ transitions and
strangeness-changing ($|\Delta S| = 1$) amplitudes mediating $b \to q {\bar q}
s$ transitions are thus related to each other, where $q$ denotes any of the
light quarks.  In the symmetry limit, the magnitudes of the two types of
amplitudes with the same flavor topology differ only in their CKM factors, and
the associated strong phases are taken to be the same.

A few questions naturally arise: (1) Do the CP asymmetries along with the rates
of the rare $B$ decays provide a coherent picture?  (2) Is it consistent with
what we have learned from other processes ({\it e.g.}, charmed $B$ decays, CP
violation in kaon decays, etc)?  (3) Is flavor symmetry breaking effects serious
for such analyses?

\section{Global fits to rare $B$ decay observables}

In recent years, several analyses
\cite{Chiang:2003pm,Chiang:2004nm,Chiang:2006ih,Chiang:2008xx} have been
performed to obtain a global fit to the measured observables in the $B^{+,0}$
decays to either two pseudoscalar mesons ($PP$) or one vector meson plus one
pseudoscalar meson ($VP$).  One could potentially extend the framework to decay
modes with two vector mesons in the final state.  However, it involves more
amplitudes of different polarizations in each flavor topology, and is beyond the
current ability to analyze due to limited available data.

\begin{figure}
\includegraphics[width=70mm]{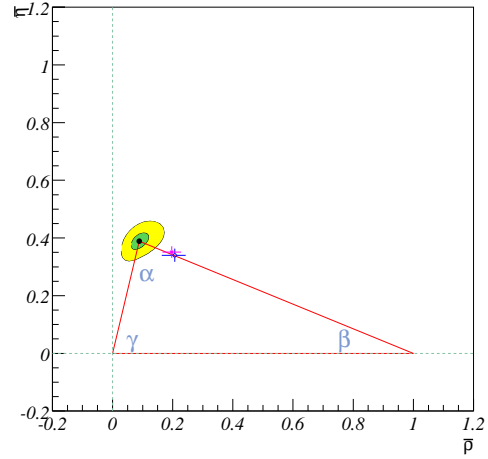}
\caption{Constraints on the $(\bar\rho,\bar\eta)$ vertex using all the $PP$ mode
  data \cite{Chiang:2006ih}.  This is consistent with other methods
  \cite{CKMfitter,UTfit} (given by the crosses) within
  errors.
  \label{fig:rhoeta-all}}
\end{figure}

In practice, an SU(3)-breaking factor is often associated to the color-allowed
tree amplitude (denoted by $T$) and color-suppressed tree amplitude (denoted by
$C$) between $\Delta S = 0$ and $|\Delta S| = 1$ transitions, but not for the
QCD penguin or electroweak penguin amplitudes (denoted by $P$ and $P_{EW}$,
respectively).  This is suggested by the factorizability of $T$ and $C$
amplitudes.  In this case, the SU(3)-breaking factor is generally taken to be
the appropriate ratio of decay constants.  For the $PP$ modes and the $VP$ modes
with the spectator quark in $B$ meson going into the vector meson, it is $f_K /
f_\pi \simeq 1.22$ \cite{PDG}.  For the $VP$ modes with the spectator quark
going into the pseudoscalar meson, it is $f_{K^*} / f_\rho \simeq 1.00$
\cite{PDG}.  Ref.~\cite{Chiang:2006ih} has found that such a breaking scheme
leads to better fits than exact symmetry for the $PP$ modes and no additional
SU(3)-breaking factor is preferred.  However, for the $VP$ modes, it is found
that the exact symmetry scheme renders better fits \cite{Chiang:2008xx}.  As a
by-product of the fitting, the UT vertex $(\bar\rho,\bar\eta)$ can be
constrained, as shown in Fig.~\ref{fig:rhoeta-all} using the $PP$ modes for
example.  It is noted that current analyses have not considered SU(3) breaking
in the strong phases.

One salient conclusion from the analyses is large relative size ($\sim 0.6$) an
strong phase ($\sim -56^\circ$) between the $C$ and $T$ amplitudes for the $PP$
decays.  This result is largely driven by the large branching ratio of the $B^0
\to \pi^0 \pi^0$ mode and the fact that $A_{CP}(B^+ \to \pi^0 K^+)$ and
$A_{CP}(B^0 \to \pi^+K^-)$ differ too much.  Interestingly, the $C_V$ and $T_V$
amplitudes also has a ratio about $0.6$ and a large relative strong phase in the
$VP$ decays.  In comparison, the ratio $|C_P/T_P|$ is only about $0.2$ - $0.3$.
Here the subscript $V$ ($P$) indicates that the spectator quark in the $B$ meson
ends up in the vector (pseudoscalar) meson in the final state.

It is worth pointing out some of the CP asymmetries predicted based upon the
best fit.  The direct CP asymmetry $A_{CP}(B^0 \to \pi^0\pi^0)$ is expected to
be at the order of $0.5$ or more, a result of comparable QCD penguin and
color-suppressed tree amplitudes and a non-trivial strong phase between them.
The $B_s \to \pi^+ K^-$ and $K^+ K^-$ modes involve the same flavor amplitudes
as the $B^0 \to \pi^+ \pi^-$ and $\pi^+ K^-$ decays, respectively.  Therefore,
they are expected to have sizable CP asymmetries due to the interference between
color-allowed tree and QCD penguin amplitudes.  The CP asymmetry of $B_s \to
\pi^+ K^-$ is predicted to be about $0.3$, which agrees well with the latest
measurement of $0.39 \pm 0.17$ by the CDF Collaboration \cite{Morello:2006pv}.
Moreover, the $B_s \to \pi^0 K_s$ decay involves the same flavor diagrams as the
$B^0 \to \pi^0 \pi^0$ mode.  Therefore, its CP asymmetries should be roughly the
same as their counterparts in $B^0 \to \pi^0 \pi^0$.

To account for the branching ratios of $B$ decays involving $\eta$ or $\eta'$ in
the final state (particularly the $\eta' K$ and $\eta K^*$ decays), one possible
solution is to have a large singlet penguin amplitude
\cite{Dighe:1995gq,Dighe:1997hm,Gronau:1995ng,Chiang:2003rb}.  Moreover, it has
a trivial strong phase with respect to the QCD penguin amplitudes in order to
produce maximal constructive or destructive interference.  However, such large
singlet amplitudes remain difficult to accommodate in the perturbative
framework \cite{Beneke:2002jn,Charng:2006zj}.

\section{The $K \pi$ puzzle}

Experimental data of the following $B \to K \pi$ modes have aroused a lot of
interest in recent years:
\begin{eqnarray}
A(B^+ \to K^0 \pi^+) &=& P' ~,
\nonumber \\
\sqrt{2} A(B^+ \to K^+ \pi^0) &=& -(P'+T'+C'+P'_{EW}) ~,
\nonumber \\
A(B^0 \to K^+\pi^-) &=& - (P'+T') ~,
\nonumber \\
\sqrt{2} A(B^0 \to K^0\pi^0) &=& P'-C'-P'_{EW} ~,
\end{eqnarray}
where the primes in the flavor amplitude decompositions denote $|\Delta S| = 1$
transitions.  A perplexing fact is first observed a few years ago by noticing
that \cite{Buras:2003yc,Yoshikawa:2003hb,Gronau:2003kj} the ratios of averaged
decay widths
\begin{eqnarray}
R_c \equiv
\frac{2 \overline\Gamma(B^+ \to K^+ \pi^0)}{\overline\Gamma(B^+ \to K^0 \pi^+)}
\mbox{ and }
R_n \equiv
\frac{\overline\Gamma(B^0\to\pi^-K^+)}
{2\overline\Gamma(B^0\to\pi^0K^0)}
\nonumber
\end{eqnarray}
are quite different.  However, they should be about the same if the $C$ and
$P_{EW}$ amplitudes are negligible, as one would na{\"i}vely expect in the SM.
This puzzle is disappearing as the two values currently become $1.12 \pm 0.07$
and $0.98 \pm 0.07$, respectively, and differ only by $1.4 \sigma$.

Nevertheless, a more serious new puzzle is emerging, for it occurs in the CP
asymmetries.  Within the SM, the difference between $A_{CP}(K^+ \pi^0)$ and
$A_{CP}(K^+ \pi^-)$ is generally expected to be small, but turns out to be
appreciably different (with a difference of $0.147 \pm 0.028$).  There are two
possible explanations for this.  One is a sizable $C'$ with a large strong phase
relative to $T'$, as found in Refs.~\cite{Chiang:2004nm,Chiang:2006ih} through
global fits in the flavor SU(3) framework.  This is also favored by the large
branching ratio of $B^0 \to \pi^0 \pi^0$.  Part of this is also justified in
pQCD analysis \cite{Li:2005kt}.  The other is a sizable new physics contribution
with a new weak phase entering through the electroweak penguin loop
\cite{Yoshikawa:2003hb,Chiang:2006ih,Hou:2006jy,Baek:2007yy,Feldmann:2008fb}.

Either of the above-mentioned solutions poses challenges for theorists.  The
former requires our better understanding of strong dynamics in the SM.  The
latter calls for more explorations in justifying the effects of physics beyond
the SM.  For example, new physics contributions are likely to affect the $VP$
counterparts of the $K \pi$ modes as well.

\section{Tests of the flavor symmetry}

In addition to performing global fits and checking their quality as mentioned
above, one can also examine the flavor symmetry principle by paying attention to
some closely related decay modes.  For example, a simple test can be done by
comparing the magnitudes of QCD penguin amplitudes, {\it i.e.}, $|P|$ obtained
from $B^0 \to K^0 K^0$ and $B^+ \to K^+ K^0$ against $|P'|$ from $B^+ \to K^0
\pi^+$.  In this case, one finds that the ratio is consistent with the ratio of
CKM factors involved in these modes, $|V_{cd} / V_{cs}|$.  This partly justifies
our use of $SU(3)_F$ as the working assumption and that $f_K / f_{\pi}$ is not
applicable to QCD penguin amplitudes.

A more sophisticated comparison can be made for the following set of decay
modes:
\begin{eqnarray}
\label{GRamp-a}
A(B^+ \to K^0 \, \pi^+) &=& P ~, \\
\label{GRamp-b}
A(B^0 \to K^+ \, \pi^-) &=&
  T \, e^{i(\delta_d+\gamma)} + P ~, \\
\label{GRamp-c}
\xi A(B_s \to K^- \, \pi^+) &=& 
  \frac1{{\tilde \lambda}} T \, e^{i(\delta_s+\gamma)}
     - {\tilde \lambda} P ~,
\end{eqnarray}
which are related by U-spin symmetry.  Here $\lambda \equiv |V_{us} / V_{ud}|
\simeq 0.2317$, and the SU(3)-breaking factor \cite{PDG,Khodjamirian:2003xk},
according to factorization,
\begin{eqnarray}
\xi \equiv 
\frac{f_K F_{B^0 \pi}(m_K^2)}{f_{\pi} F_{B_sK}(m_{\pi}^2)}
\frac{m_{B^0}^2 - m_{\pi}^2}{m_{B_s}^2 - m_K^2}
= 0.97^{+0.09}_{-0.11}
\label{eq:xi}
\end{eqnarray}
corresponds to almost exact symmetry.

It has been proposed \cite{Gronau:2000md,Chiang:2000za} to extract the weak
phase $\gamma$ from the branching ratios and CP asymmetries of these modes,
assuming the same relative strong phase $\delta_d = \delta_s$.  Given the fact
that $\gamma$ has been constrained using other methods ({\it e.g.}, $DK$ modes)
and that the last mode in Eq.~(\ref{GRamp-c}) is not measured until recently by
the CDF Collaboration (see Table~\ref{tab:Kpis}), one can turn the argument
around to test the flavor symmetry assumption.  Moreover, its branching ratio
still has a large uncertainty.  Therefore, its central value may change before
it completely settles down.

\begin{table}[h]
\caption{Current data of several $B \to K \pi$ decays.  Branching ratios are
  quoted in units of $10^{-6}$.
  \label{tab:Kpis}}
\begin{tabular}{lrc}
\hline\hline
Observable & Exp.\ Value & Ref.\ \\
\hline
$BR(B^+ \to K^0 \pi^+)$ & $23.1 \pm 1.0$ & \cite{HFAG} \\
$BR(B^0 \to K^+ \pi^-)$ & $19.4 \pm 0.6$ & \cite{HFAG} \\
$A_{CP}(B^0 \to K^+ \pi^-)$ & $-0.097 \pm 0.012$ & \cite{HFAG} \\
$BR(B_s \to K^- \pi^+)$ & $5.27 \pm 1.17$ & \cite{Morello:2006pv} \\
$A_{CP}(B_s \to K^- \pi^+)$ & $0.39 \pm 0.17$ & \cite{Morello:2006pv} \\
\hline\hline
\end{tabular}
\end{table}

It is useful to consider the following four quantities:
\begin{eqnarray}
&&
R_d = 1 + r^2 + 2 r \cos\gamma \cos\delta_d = 0.899 \pm 0.048 ~,
\nonumber \\
&&
\xi^2 R_s = 
{\tilde \lambda}^2 + \left( \frac{r}{\tilde \lambda} \right)^2
- 2 r \cos\gamma \cos\delta_s = 0.260 \pm 0.059 ~,
\nonumber \\
&&
R_d A_{CP}(B^0 \to K^+ \pi^-) = 2 r \sin\gamma \sin\delta_d 
\nonumber \\
&& \qquad 
= 0.087 \pm 0.012 ~,
\nonumber \\
&&
\xi^2 R_s A_{CP}(B_s \to K^- \pi^+) = -2 r \sin\gamma \sin\delta_s 
\nonumber \\
&& \qquad
= -0.101 \pm 0.050 ~.
\label{eq:obs}
\end{eqnarray}
Here $r$ denotes the ratio between $|T|$ and $|P|$.  The last two in
Eqs.~(\ref{eq:obs}) imply a simple relation between the strong phases
\begin{eqnarray}
\frac{\sin\delta_d}{\sin\delta_s} = 0.96 \pm 0.54 ~.
\end{eqnarray}
Ref.~\cite{Chiang:2008vc} notices that one cannot obtain a solution to
Eqs.~(\ref{eq:obs}) by assuming $\delta_d = \delta_s$.

\begin{figure}[h]
\includegraphics[width=70mm]{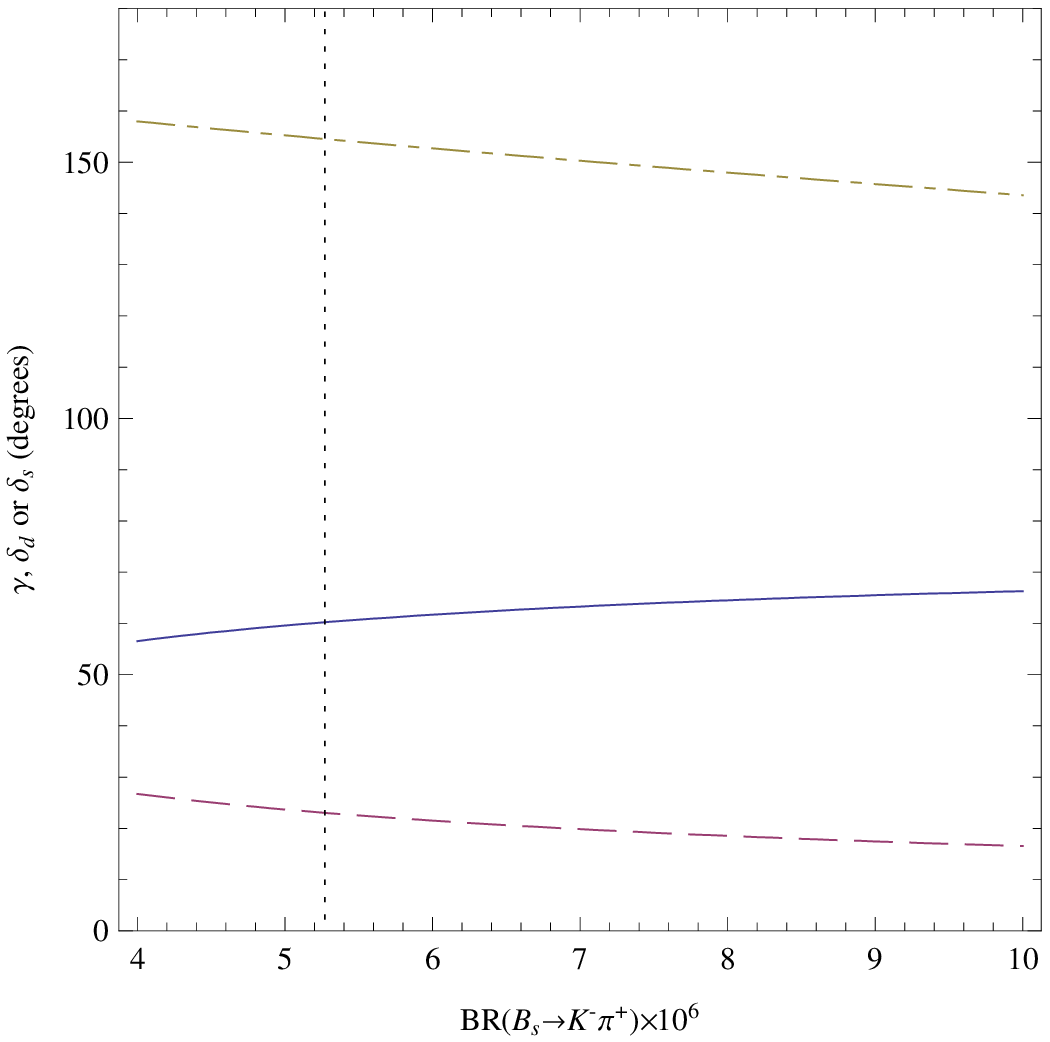} \\
\includegraphics[width=70mm]{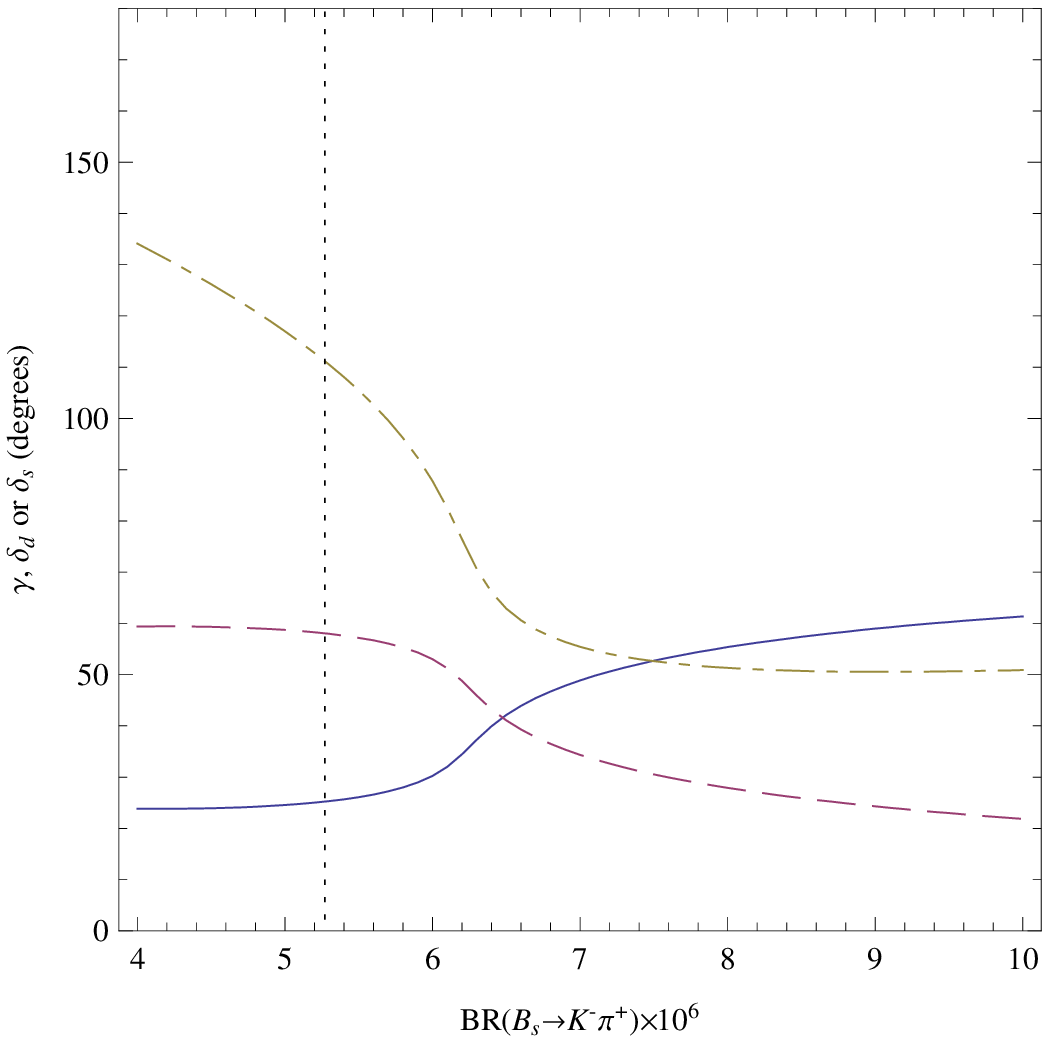}
\caption{Behavior of solutions as a function of $BR(B_s \to K^- \pi^+)$
  \cite{Chiang:2008vc}.  The solid, dashed, and dash-dotted curves represent
  $\gamma$, $\delta_d$, and $\delta_s$, respectively.  The vertical dotted line
  indicates the current value of $BR(B_s \to K^- \pi^+)$.
  \label{fig:sols}}
\end{figure}

When the equality condition on strong phases is relaxed, two sets of solutions
are obtained from the four equations in (\ref{eq:obs}) \cite{Chiang:2008vc}, as
shown in Fig.~\ref{fig:sols}.  One set (upper plot) has very different
$\delta_d$ and $\delta_s$, suggesting large SU(3) breaking in the strong phases.
The other set (lower plot) gives reasonable strong phases and weak phase
$\gamma$ provided $BR(B_s \to K^- \pi^+)$ is larger than the current value by
about 40\%.  This is possible if either recent evaluations of $b$ quark
fragmentation \cite{Aaltonen:2008zd} had overestimated the fraction of $b$
quarks ending up as $B_s$ or the SU(3) breaking factor $\xi$ is $20\%$ larger
than that given in (\ref{eq:xi}).

\section{Extraction of $\gamma$ from charmless modes}

Several methods have been proposed to determine the weak phase $\gamma$ using
the direct CP asymmetries resulted from interference between different
amplitudes in $B \to D^{(*)} K$ decays
\cite{Gronau:1991dp,Gronau:1990ra,Atwood:2000ck}.  However, such early proposals
are not completely free from hadronic uncertainties.  Recently, a Dalitz plot
analysis is used to simultaneously determine $\gamma$ and other hadronic
parameters in the problem \cite{Giri:2003ty,Poluektov:2004mf}.

As a complementary means, Refs.~\cite{Neubert:1998pt,Gronau:2003kj} suggest to
make use of the rates and asymmetries of charmless $B \to K \pi$ modes.  With
the input of the color-allowed tree amplitude from $B \to \pi \ell \nu$ decay,
one can obtain a constraint on $\gamma$.  As data in the charmless $VP$ modes
become available, it is possible to use the $K^* \pi$ and $\rho K$ final states
of the $B$ decays to constrain $\gamma$ \cite{Sun:2003wn,Chiang:2005kz}.

\begin{table}
\caption{Current rate and asymmetry data of some $B \to K^* \pi$ and $\rho K$
  modes.
  \label{tab:rhoK}}
\begin{tabular}{llccc}
\hline\hline
 & Mode & Amplitudes & BR ($\times 10^{-6}$) & $A_{CP}$ \\ 
\hline
$B^+ \to$
    & $K^{*0} \pi^+$
        & $P'_P$
        & $10.7 \pm 0.8$  
        & $-0.085 \pm 0.057$ \\
    & $\rho^+ K^0$
        & $P'_V$
        & $8.0 \pm 1.5$ 
        & $0.12 \pm 0.17$ \\
\hline
$B^0 \to$
    & $K^{*+} \pi^-$
        & $-(P'_P + T'_P)$ 
        & $9.8 \pm 1.1$ 
        & $-0.05 \pm 0.14$ \\
    & $\rho^- K^+$
        & $-(P'_V + T'_V)$
        & $15.3 \pm 3.6$ 
        & $0.22 \pm 0.23$ \\
\hline\hline
\end{tabular}
\end{table}

As shown in Table~\ref{tab:rhoK}, the CP asymmetries of all the listed modes are
consistent with zero.  In particular, the two neutral $B$ decays imply that the
strong phases between the color-allowed tree and QCD penguin amplitudes are
trivial.  We consider the following four quantities:
\begin{eqnarray}
R(K^* \pi)
&\equiv& \frac{\overline\Gamma(K^{*+} \pi^-)}{\overline\Gamma(K^{*0} \pi^+)}
= 1 - 2 r_1 \cos\delta_P \cos\gamma + r_1^2
\nonumber \\
&=& 0.99 \pm 0.13 ~,
\nonumber \\
{\cal A}_{CP}^{K^{*+} \pi^-}
&=& - 2 r_1 \sin\delta_P \sin\gamma / R(K^{*+} \pi^-)
\nonumber \\
&=& -0.05 \pm 0.14 ~.
\nonumber \\
R(\rho^- K^+)
&\equiv& \frac{\overline\Gamma(\rho^- K^{+})}{\overline\Gamma(\rho^+ K^{0})}
= 1 + 2 r_2 \cos\delta_V \cos\gamma + r_2^2
\nonumber \\
&=& 2.06 \pm 0.61 ~,
\nonumber \\
{\cal A}_{CP}^{\rho^- K^+}
&=& 2 r_2 \sin\delta_V \sin\gamma / R(\rho^- K^+) 
\nonumber \\
&=& 0.22 \pm 0.23 ~,
\label{eq:fourobs}
\end{eqnarray}
where $r_1 \equiv |T'_P / P'_P|$, $r_2 \equiv |T'_V / P'_V|$, and $\delta_P$ and
$\delta_V$ are the corresponding relative strong phases.  Instead of treating
$r_1$ and $r_2$ as independent parameters, one may employ the factorization
assumption to obtain
\begin{eqnarray}
\frac{r_2}{r_1}
= \frac{f_K A_0^{B \rho}(m_K^2)}{f_{K^*} F_1^{B \pi}(m_{K^*}^2)}
\left| \frac{P'_P}{P'_V} \right|
= 0.6 - 1.1 ~,
\end{eqnarray}
where the uncertainty mainly comes from the form factors.  In this case, there
are now only four parameters for the four observables in
Eqs.~(\ref{eq:fourobs}).  Solving them gives $\gamma = (65^{+10}_{-8})^\circ$
for $r_2/r_1 = 0.6$ and $(68^{+9}_{-7})^\circ$ for $r_2/r_1 = 1.1$.  This result
is consistent with other methods \cite{CKMfitter,UTfit}.

\section{Summary}

We review the importance of CP asymmetries in $B$ meson decays.  The flavor
SU(3) symmetry assumption is employed to analyze charmless two-body modes in a
global way.  The direct CP asymmetries provide useful information on both weak
and strong phases.  Currently, there are six direct CP asymmetry observables
deviating from zero at $3\sigma$ level or more.  Sizable relative sizes and
strong phases are observed between the color-allowed and color-suppressed tree
amplitudes from current data.  The puzzle in the CP asymmetry pattern of $B \to
K \pi$ decays can be explained by such a color-suppressed amplitude or some new
physics effects with a new weak phase in the electroweak penguin amplitude.  The
large color-suppressed amplitude explanation requires better understanding of
the strong dynamics.  The same comment applies to the singlet penguin amplitudes
too.

We examine the flavor symmetry principle by scrutinizing a set of $B_{u,d,s} \to
K \pi$ decays.  Current data indicate unexpectedly large SU(3) breaking in the
strong phases.  This may eventually go away if $BR(B_s \to K^- \pi^+)$ turns out
to be $40\%$ larger or the symmetry breaking in amplitude sizes is $20\%$
larger.  A definite conclusion on this issue relies on more precise experimental
measurements.  In addition to the $B \to D^{(*)} K$ modes, it is useful to
constrain $\gamma$ using charmless decay modes as well.  The rates and CP
asymmetries of some $K^* \pi$ and $\rho K$ modes can be used to determine
$\gamma$ with only a mild assumption of factorization for the tree amplitudes.

\begin{acknowledgments}
  The author is grateful to many of his collaborators on various aspects of the
  topic.  This work is supported in part by the National Science Council of
  Taiwan, ROC under Grant No. NSC 96-2112-M-008-001.
\end{acknowledgments}

\end{document}